\documentclass[aps,prl,reprint,showpacs,superscriptaddress]{revtex4-1}
\usepackage{graphicx}
\usepackage[utf8]{inputenc}
\usepackage[english]{babel}
\usepackage{color}
\usepackage{soul}
\begin{document}

\title{Universal Depinning Transition of Domain Walls in Ultrathin Ferromagnets\\}

\author{R. Diaz Pardo}
\affiliation{Laboratoire de Physique des Solides, Universit\'e Paris-Sud, CNRS, UMR8502, 91405 Orsay, France.}
\author{W. Savero Torres}
\affiliation{Laboratoire de Physique des Solides, Universit\'e Paris-Sud, CNRS, UMR8502, 91405 Orsay, France.}
\author{A. B. Kolton}
\affiliation{CONICET, Centro At\'omico Bariloche, 8400 San Carlos de Bariloche, R\'{\i}o Negro, Argentina.}
\author{S. Bustingorry}
\affiliation{CONICET, Centro At\'omico Bariloche, 8400 San Carlos de Bariloche, R\'{\i}o Negro, Argentina.}
\author{V. Jeudy}
\email[]{vincent.jeudy@u-psud.fr}
\affiliation{Laboratoire de Physique des Solides, Universit\'e Paris-Sud, CNRS, UMR8502, 91405 Orsay, France.}

\date{\today}

\begin{abstract}
We present a quantitative and comparative study of magnetic field driven domain wall depinning transition in different ferromagnetic ultrathin films over a wide range of temperature.
We reveal a universal scaling function accounting for both drive and thermal effects on the depinning transition, including critical exponents. 
The consistent description we obtain for both the depinning and subthreshold thermally activated creep motion should shed light on the universal glassy dynamics of thermally fluctuating elastic objects pinned by disordered energy landscapes.

\end{abstract}


\maketitle
The depinning transition from a pinned to a sliding state upon increasing a driving force, is a nonequilibrium phenomenon
observed in extremely diverse systems, ranging from
fractures~\cite{ponson_PRL_2009,laurson_natcomm_2013},  
charge density waves~\cite{fisher_PRB_85_charge_density_waves}, 
vortex glasses in superconductors~{\cite{blatter_vortex_review,review_ledoussal_2010}},
ferroelectric~\cite{Jo_PRL_2009_ferroelectric,Li_JPD_2005_depinning_ferroelectric}
and ferromagnetic materials~\cite{zapperi_PRB_1998,metaxas_PRL_07_depinning_thermal_rounding,bustingorry_PRB_12_thermal_rounding,Durin_prl2016}
to wetting~\cite{ledoussal_epl2009}, reaction~\cite{Atis_prl2015} and 
cell migration advancing fronts~\cite{Chepizhko_pnas2016}.
At zero temperature, a pinned elastic system presents, upon increasing the driving force $f$,  
a depinning threshold $f_d$ separating the zero velocity state for $f<f_d$ from a finite velocity regime for $f>f_d$.
At finite temperature and below the depinning threshold, the thermal fluctuations result in a so-called "creep" motion over effective pinning barriers.
In this dynamical regime, the velocity follows an Arrhenius law $v\sim \exp(-\Delta E/k_B T)$ where $k_BT$ is thermal activation (and $k_B$ the Boltzmann constant). 
%
Close to zero drive ($f \rightarrow 0$), phenomenological scaling theory ~\cite{feigelman1989,nattermann1990} and functional renormalization group~\cite{chauve2000} calculations for an elastic line moving in a random pinning disorder predict the effective pinning barriers present a universal power law variation $\Delta E\sim f^{-\mu}$ with a critical exponent value $\mu=1/4$ which was observed in a real system~\cite{lemerle_PRL_1998_domainwall_creep}. More recently, it was shown experimentally that the whole thermally activated dynamical regimes up to the depinning threshold can be universal and controlled by a unique pinning energy barrier function~\cite{jeudy_PRL_2016_energy_barrier}.

For the depinning transition, the situation is much less clear since the pinning barriers vanish and their height becomes comparable to thermal activation. 
Statistical physics approaches predict universal scaling functions of the driving force and temperature for the velocity near depinning
~\cite{
middleton_PRB_92_CDW_thermal_exponent,
Roters_PRE_1999_depinning,chauve2000,
ledoussal_PRB_2002_depinning_transition,Rosso_PRE_2003,
bustingorry_epl_2008_thermal_rounding,bustingorry_PRE_12_thermal_rounding}.
%
%
Above the depinning threshold, it is well established that the velocity displays the critical behavior $v \sim (f-f_d)^\beta$,
$\beta$ being the depinning exponent~\cite{fisher_PRB_85_charge_density_waves,middleton_PRB_92_CDW_thermal_exponent,Ferrero_PRE_2013_depinning}. At finite temperature, the combined effects of drive and thermal noise produce a ``thermal rounding'' of the velocity-force characteristics. The velocity is usually described by $v \sim T^\psi$ at the threshold ($f=f_d$), with $\psi$ the thermal rounding exponent
~\cite{middleton_PRB_92_CDW_thermal_exponent,bustingorry_epl_2008_thermal_rounding,bustingorry_PRE_12_thermal_rounding}.
Just above the depinning threshold ($f \gtrsim f_d$), the combined effects of the drive and thermal noise are conjectured to be caught by $v \sim T^\psi g[T^{-\psi}(f-f_d)^\beta]$, where $g$ is a universal scaling function~\cite{fisher_PRB_85_charge_density_waves,middleton_PRB_92_CDW_thermal_exponent,ledoussal_PRB_2002_depinning_transition,Rosso_PRE_2003,bustingorry_epl_2008_thermal_rounding,bustingorry_PRE_12_thermal_rounding}. 
This kind of response, phenomenologically predicted by exploiting the analogy with equilibrium phase transitions, 
is supported by numerical simulations of a driven elastic string~\cite{bustingorry_epl_2008_thermal_rounding,bustingorry_PRE_12_thermal_rounding}, the random-field Ising model 
\cite{Roters_PRE_1999_depinning} 
and charge density waves models \cite{middleton_PRB_92_CDW_thermal_exponent}
but still remains elusive to a rigorous FRG treatment ~\cite{chauve2000}.
Experimentally on the other hand, it has been extremely challenging to go beyond critical exponent analysis~
\cite{ponson_PRL_2009,Jo_PRL_2009_ferroelectric,Li_JPD_2005_depinning_ferroelectric,zapperi_PRB_1998}. Investigations of the depinning transition is complicated by the thermal rounding of the velocity-force characteristics~\cite{bustingorry_epl_2008_thermal_rounding,bustingorry_PRE_12_thermal_rounding}
which impedes a straightforward determination of the depinning threshold and consequently a clear distinction between material dependent and universal behaviors. 
%
Therefore, assessing the very existence of a universal scaling function of a reduced force and temperature remains an important open issue.

In this work, we address the question of the universality of the depinning transition in presence of drive and thermal fluctuations, 
going beyond the determination of critical exponents
and we evidence a universal function which captures both the temperature and external drive scaling properties.
Strategically, we have chosen to study the motion of domain walls driven by magnetic field in an ultrathin film with perpendicular anisotropy~\cite{metaxas_PRL_07_depinning_thermal_rounding,gorchon_PRL_2014}. In this system, the magnetic field corresponds to the driving force and the domain walls to elastic one-dimensional lines moving in a two-dimensional medium. The sub-threshold creep universal behavior is well understood, 
so that the non-universal material dependent parameters can be treated consistently.   
The effective energy barrier~\cite{jeudy_PRL_2016_energy_barrier} $\Delta E \sim (H/H_d)^{-\mu}-1$, with a creep exponent $\mu=1/4$~\cite{chauve2000, lemerle_PRL_1998_domainwall_creep,metaxas_PRL_07_depinning_thermal_rounding} describes the whole sub-threshold regime up to the depinning threshold. The depinning thresholds as well as the other material dependent parameters are non-ambiguously determined from the methods developped in Ref.~\cite{jeudy_PRL_2016_energy_barrier} and are then used to analyze the depinning transition.
 At and above threshold, 
we use the depinning critical exponents $\beta = 0.25$ \cite{Ferrero_PRE_2013_depinning} and $\psi=0.15$ \cite{bustingorry_epl_2008_thermal_rounding,bustingorry_PRE_12_thermal_rounding} which are deduced from recent numerical simulations and are compatible with experimental findings~\cite{gorchon_PRL_2014}.

\begin{figure}[ht]
\centering
\includegraphics[width= 7.6 cm,height= 6.2 cm]{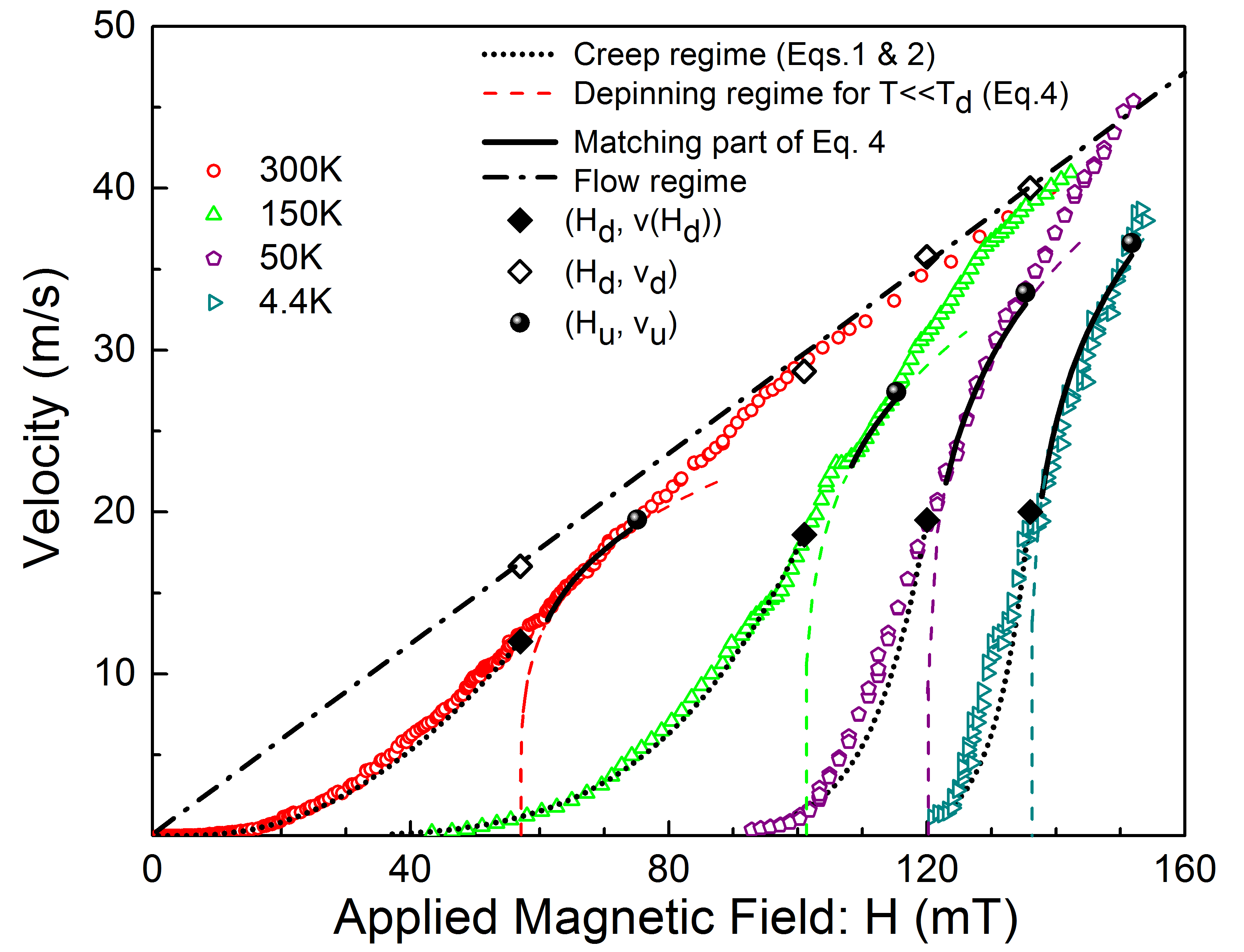}

\caption{The velocities measured for different temperatures. 
%
The {\it creep regime} $[0<H<H_d(T)]$ is highlighted by the black dotted curves corresponding to fits of Eq.~(\ref{eq:v-creep}) and ends for velocities $v(H_d(T),T)$ indicated by black diamond symbols.
The {\it depinning regime} extends from the depinning threshold $H=H_d(T)$ to the universality limit $H_u(T)$ (indicated by black solid spheres). The depinning velocities $v_T(H_d(T),T)$ deduced form Eq. \ref{eq:depinning_T} correspond to empty diamonds.
The dashed curves are the predictions of Eq.~(\ref{eq:depinning_H}) for a unique value of $x_0=v_T/v_H=0.65$ and match with {part of all velocity curves as indicated by black solid segments}.  
The {\it linear flow regime}, indicated by a straight dashed-dot line, is reached for sufficiently high drive (and only for $T> 50 K$). 
%
}
\label{fig:1}
 \end{figure}

The sample is a Pt(3.5~nm)/Co(0.45~nm)/Pt(4.5~nm) ultrathin film grown by sputtering on an etched Si/SiO$_2$ substrate\cite{metaxas_PRL_07_depinning_thermal_rounding}. 
In order to test the universal functions, the domain wall dynamics is analyzed over large ranges of magnetic field (0-160mT) using a small coil placed on the film ($\sim 100$ turns and $\sim $ 1mm in diameter) and of temperature ($4.4-300$K) using an optical He-flow cryostat.
Polar magneto-optical Kerr effect microscopy (PMOKE) is used to observe domain walls. Their motion is produced by magnetic field pulses applied perpendicular to the film. The velocity curves shown in Fig. \ref{fig:1} result from a sliding average over 5 points used to smooth raw data. The measured velocity corresponds to the ratio between the domain wall displacements and the pulse duration.  The data are then compared to results published in the literature for Au/Co/Au (Ref. \onlinecite{kirilyuk_JMMM_97_AuCoAu}) and CoFeB (Ref. \onlinecite{Burrows_APL_13_CoFeB}) ultrathin films, thus allowing to explore different strengths of pinning disorder (see supplementary material \onlinecite{supplementary_material}).

The velocity curves obtained for Pt/Co/Pt ultrathin films over a large temperature range (see Fig. \ref{fig:1}) reveal different dynamic regimes.
An inflection point in the velocity-field curve (see Fig. \ref{fig:1}) separates the low drive $[0<H<H_d(T)]$ {\it creep regime}, where the velocity varies several orders of magnitude over a relatively narrow applied magnetic field range (see Fig. \ref{fig:1}(b)), from the {\it depinning regime} observed for $[H \gtrsim H_d(T)]$. At sufficient large drive, domain walls follow the {\it flow regime} characterized by a linear variation of the velocity (see Fig. \ref{fig:1}).

In order to discuss those different regimes on a quantitative basis, we propose a unified description of the glassy domain wall dynamics. 
For the whole creep regime, the domain wall velocity is described~\cite{jeudy_PRL_2016_energy_barrier} by the following expression:

\begin{equation}
 \label{eq:v-creep}
 v(H,T) = v(H_d,T) \exp\left( \frac{-\Delta E}{k_B T} \right)
\end{equation}
with an effective pinning energy barrier
\begin{equation}
\label{eq:E-creep}
 \Delta E = k_B T_d \left[ \left( \frac{H}{H_d}\right)^{-\mu} - 1 \right],
\end{equation}
where $H_d$ and $v(H_d,T)$ are the coordinates of the upper boundary of the creep regime ($\Delta E \rightarrow 0$) 
and $k_B T_d$ is a characteristic energy scale measuring the pinning strength.
The parameters $v(H_d(T),T)$, $H_d(T)$ and $T_d(T)$ were determined from the creep motion for each temperature using the method developped in Ref. \onlinecite{jeudy_PRL_2016_energy_barrier} (see Ref. \onlinecite{supplementary_material} for details and the values). As shown in Fig. 1, the fit of Eqs.~\ref{eq:v-creep} and~\ref{eq:E-creep} present a good agreement over the whole creep regime.

For the depinning transition, the asymptotic power laws can be written as
\begin{equation}
\label{eq:depinning_T}
 v(H_d,T) =  v_T \left( \frac{T}{T_d} \right)^{\psi}.
\end{equation} 
reflecting the variation with temperature at the threshold ($H=H_d$) and
\begin{equation}
v(H,T=0K)=v_H \left( \frac{H-H_d}{H_d}\right)^{\beta},
\label{eq:depinning_H}
\end{equation}
corresponding to the variation with magnetic field at zero temperature ($T=0K$). Here, $v_T$ and $v_H$ are depinning velocities.
Following the more general scaling hypothesis \cite{stanley_book_1971}, the velocity can be described as a generalized homogeneous function~\cite{privman_book_1991,bustingorry_PRE_12_thermal_rounding}, which implies the following scaling form:
\begin{equation}
y= g\left(\frac{x}{x_0} \right)
\label{eq:g_function}
\end{equation}
where we have defined the scaled dimensionless variables
$y=(v/v_T)(T/T_d)^{-\psi}$ and $x=[(H-H_d)/H_d]^\beta (T/T_d)^{-\psi}$, and where $x_0=v_T/v_H$ is the amplitude ratio~\cite{privman_book_1991} of the depinning velocities. As the depinning velocities $v_T$ and $v_H$ describe the same physics at the Larkin lenghscale \cite{supplementary_material}, the value of $x_0$ should be close to 1.
In this model, the parameters $v_T$, $v_H$, $H_d$ and $T_d$ are non-universal material and temperature dependent.
Since $v_T$ and $v_H$ act here as purely metric factors~\cite{privman_book_1991}, {\it a priori} the function $g$ is expected to be universal within a given class of universality and $x_0$ to be a non-universal parameter.
The shape of the function $g$ should reflect the asymptotic behavior described by Eqs. \ref{eq:depinning_T} and \ref{eq:depinning_H} and should enlight the respective contribution of drive and temperature on the velocity curves. For $x \ll x_0$ (i.e., $H-H_d \ll H_d(T_d/T)^{-\psi/\beta }$), we should have  $g(x/x_0) \rightarrow 1$ which corresponds to Eq. \ref{eq:depinning_T}. Therefore, the velocity curves (shown in Fig. \ref{fig:1}) are expected to present important thermal effects only sufficiently close to the depinning threshold. On the opposite, for $x \gg x_0$ (i.e., $H-H_d \gg H_d(T_d/T)^{-\psi/\beta }$), we expect $g(x/x_0) \rightarrow x/x_0$ in agreement with the zero temperature asymptotic limit (Eq. \ref{eq:depinning_H}). As a consequence, sufficiently far from the depinning threshold, the velocity should present a temperature independent variation with magnetic field.

%
%
%

\begin{figure}[ht]
 \centering
\includegraphics[width=7.6cm,height=4.2cm]{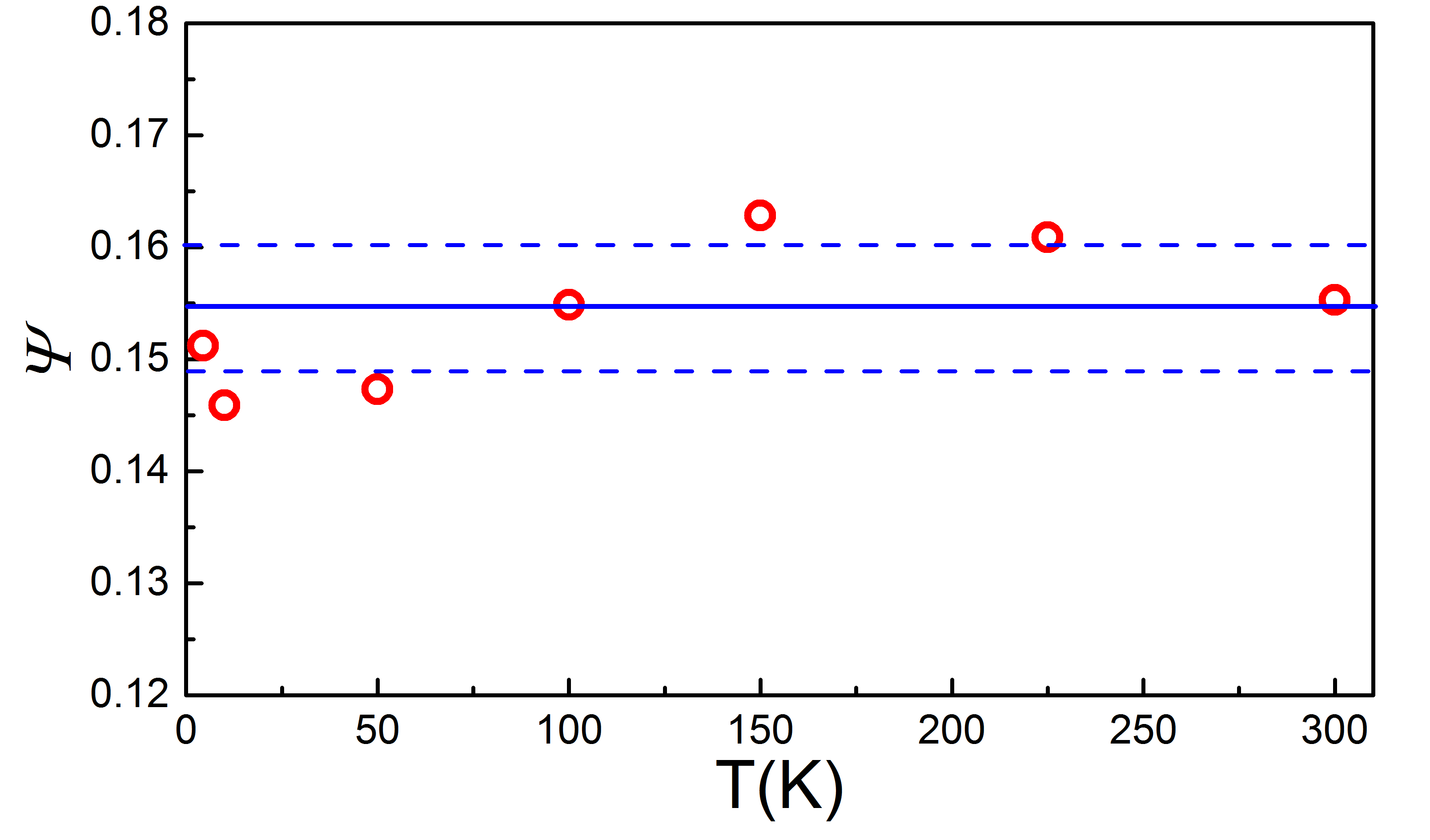}
 \caption{Depinning critical exponent $\psi$ deduced from experimental curves of Fig. \ref{fig:1} and shown as a function of temperature.%
}
\label{fig:2}
 \end{figure}

Let us now discuss the experimental results. At the depinning threshold $(H=H_d)$, Eq.~(\ref{eq:depinning_T}) (with $\psi=0.15$) allows to determine the value of the depinning velocity $v_T$ for different temperatures. As it can be seen in Fig. \ref{fig:1}, $v_T$ presents systematically a good agreement with the velocity corresponding to the linear flow regime. This suggests the depinning velocity $v_T$ to correspond to the velocity  DW would reach in absence of pinning. Moreover, assuming $v_T=mH_d$, where $m$ is the slope of the flow regime, and inverting Eq.~(\ref{eq:depinning_T}), we can estimate the critical exponent $\psi$. As shown in Fig.\ref{fig:2}, $\psi$ is found to be temperature independent and equal to 0.154 $\pm$ 0.006. The  good agreement with previous experimental findings \cite{gorchon_PRL_2014} and with numerical predictions \cite{bustingorry_epl_2008_thermal_rounding} is a first signature of the universality of the depinning transition.

Above the depinning threshold, since the temperature ratio $T_d/T$ is large ($>10$) (see \cite{supplementary_material}), 
part of velocity curves should follow the predictions of Eq.~(\ref{eq:depinning_H}) with $\beta=0.25$.
In order to test this hypothesis, the values of $v_H$ were adjusted so that the predictions present the largest matching with the velocity curves (see Fig. \ref{fig:1}). As expected, a good agreement is only obtained over a limited field range starting slightly above $H_d$ ($H-H_d>1-7mT$ for $T=4.4-300K$) due to thermal effect and whose upper boundary $H_u(T)$ defines the limit of the universal depinning transition.
Above $H_u(T)$, DWs should follow a non-universal crossover to the linear flow regime observed at larger drive.

\begin{figure}[ht]
 \centering
\includegraphics[width=7.6cm,height=4.47cm]{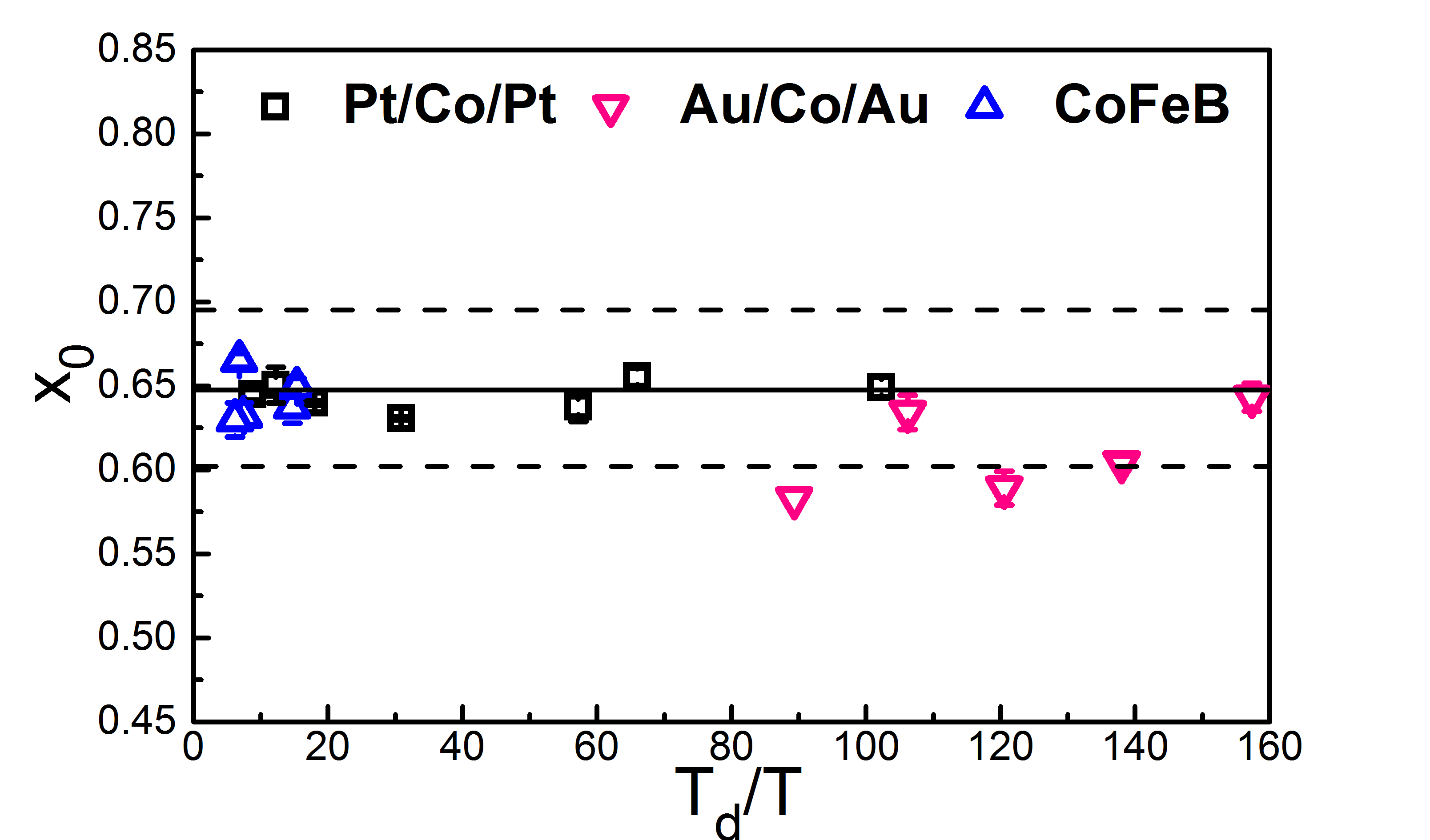}
 \caption{Universal velocity ratio, $x_0=v_T/v_H$, as a function of the reduced temperature $T_d/T$ for different ferromagnetic materials. The solid and dashed lines indicate the average value and standard deviation deduced from a fit of the curve in Fig. \ref{fig:3} (see text), respectively.%
} 
\label{fig:2b}
 \end{figure}

We now investigate the universality of the whole depinning transition $(H_d<H<H_u)$ beyond power laws and critical exponents. Let us first discuss the variations of the ratio $x_0=v_T/v_H$ with the reduced temperature $T_d/T$. Surprisingly, $x_0$ is found to be temperature independent ($x_0$ =$0.64 \pm 0.02$)
as shown in Fig. \ref{fig:2b}. As direct consequence, a unique value of $x_0$ allows the predictions of Eq.~(\ref{eq:depinning_H}) to describe the full set of velocity curves 
(see Fig. \ref{fig:1}). 
Moreover, an identical analysis can be performed for other ferromagnetic ultrathin films. As shown in Fig. \ref{fig:2b}, the obtained results present a particularly good agreement for CoFeB\cite{Burrows_APL_13_CoFeB}($x_0$=$0.64 \pm 0.02$) and a slightly lower values $x_0$ (=$0.62 \pm 0.02$) for Au/Co/Au\cite{kirilyuk_JMMM_97_AuCoAu}.
%
 Therefore as the mean values vary by less than 5$\%$ over wide explored range of reduced temperature ($10<T_d/T<170$), $x_0$ can be reasonably considered as material and temperature independent, which was {\it a priori} not predicted. This result implies that only three material and temperature dependent parameters ($v_T=x_0v_H$, $H_d$ and $T_d$) are sufficient to describe the whole DW glassy dynamics (see    
Ref. \cite{supplementary_material}).

\begin{figure}[ht]
 \centering
\includegraphics[width=7.6cm,height=6.2cm]{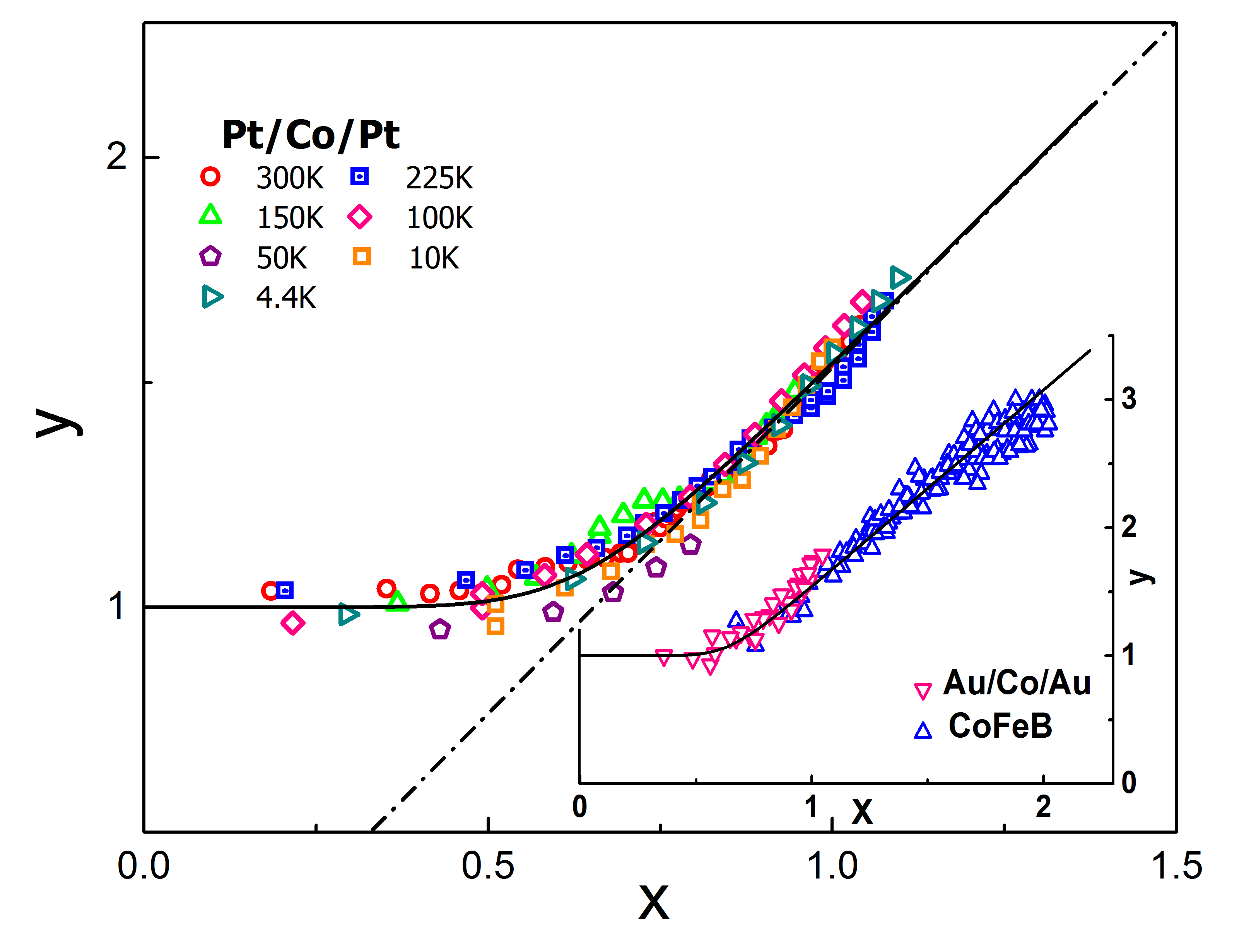}
\caption{Universal depinning scaling function obtained using the scaled domain wall velocity $y=(v/v_T)(T/T_d)^{-\psi}$ as a function of the scaled field $x=[(H-H_d)/H_d]^\beta (T/T_d)^{-\psi}$. The data points only correspond to universal depinning transition ($H_d<H<H_u$).
The solid curve (also shown the inset) is an empirical function $g(x/x_0)$ describing the data (see text).
The dashed-dotted straight line is the linear asymptotic limit of function $g(x/x_0)$. \textit{Inset:} Scaled domain wall velocity $y$ as a function of scaled field $x$ for Au/Co/Au\cite{kirilyuk_JMMM_97_AuCoAu}, and for CoFeB\cite{Burrows_APL_13_CoFeB}.%
}
\label{fig:3}
 \end{figure}

Moreover, following Eq.~(\ref{eq:g_function}) both magnetic field and temperature effects on DW dynamics should be described by a single function $g$.
%
Figure~\ref{fig:3} shows experimental values for the scaled velocity $y=(v/v_T)(T/T_d)^{-\psi}$ as a function of the scaled driving field $x=[(H-H_d)/H_d]^\beta (T/T_d)^{-\psi}$. All the velocity curves collapse onto a single master curve thus indicating that $x$ and $y$ are appropriate reduced variables to describe the depinning transition and the existence of a universal function $g$. 
The expected linear asymptotic behavior $g(x/x_0) \rightarrow x/x_0$ (derived from Eq.~(\ref{eq:depinning_H})) and corresponding to the low temperature limit ($T_d/T \gg 1$) is observed for $x > 0.7$. For $x<0.5$, the scaled field $y$ remains almost constant and data extrapolate to $y=1$ for $x \rightarrow 0$ as derived from Eq.~(\ref{eq:depinning_T}) (for $H \rightarrow H_d$). 
 Therefore the universal $g$-function essentially displays two linear asymptotical behaviors and a narrow crossover region ($0.5<x<0.7$), which is in qualitative agreement with predictions deduced from numerical simulation in Ref. \cite{bustingorry_PRE_12_thermal_rounding} which found $x_0 \sim 1$. 
 A rather accurate empirical description of data (see the solid lines in Fig.\ref{fig:3}) is given by: $g(x)=\left[1+\left( \frac{x}{x_0} \right)^n \right]^{1/n}$,
where $n$ reflects the width of the crossover~\cite{note}, with a best fit obtained for $x_0=0.65 \pm 0.04$ and $n=8.7 \pm 0.4$.  
 Moreover, this law is also found to be relevant for other magnetic materials (Au/Co/Au\cite{kirilyuk_JMMM_97_AuCoAu}, and CoFeB\cite{Burrows_APL_13_CoFeB}) (see  the inset of Fig. \ref{fig:3}), bringing further evidence that the depinning transition is described by a unique universal function.

In conclusion, the depinning transition of domain walls driven by magnetic field in ultrathin films has been shown to present a universal behavior. The latter is characterized by a universal function of the rescaled field, temperature and velocity whose shape governs the domain wall velocity including its asymtoptic scaling law behaviors. 
Moreover, the proposed phenomenological model and a single set of material and temperature dependent 
parameters ($H_d$, $T_d$ and $v_T$) are found to allow a self-consistent analysis of both the depinning transition and the sub-threshold thermally activated creep regime.
Our study is thus relevant for describing, in an unified way,  
driving and thermal effects in the glassy dynamics of driven disordered 
elastic systems. 

\begin{acknowledgments}
We wish to thank A. Mougin, A. Thiaville, and the PAREDOM collaboration in Bariloche for fruitful discussions. S. B., J. G., and V. J. acknowledge support by the French-Argentina project ECOS-Sud num.A12E03. This work was also partly supported by the french projects DIM CNano IdF (Region Ile-de-France) and the Labex NanoSaclay, reference: ANR-10-LABX-0035. R.D.P. thanks the Mexican council CONACyT for the PhD fellowship n0: 449563. S. B. and A. B. K. acknowledge partial support from Project PIP11220120100250CO (CONICET).

\end{acknowledgments}

\bibliography{ref_depinning_AK_2}

%
%

\end{document}